\documentclass[twocolumn,twocolappendix]{aastex7}
\usepackage[whole]{bxcjkjatype}
\usepackage{amsmath}
\usepackage[dvipdfmx]{xcolor}
\received{2025 August 28}
\revised{2026 January 13}
\accepted{2026 January 13}
\submitjournal{ApJ}

\begin{document}
\title{Ring-Gap Structures in the Class I Circumstellar Disk of CrA~IRS~2 Associated with Magnetic Flux-Driven Bubble}

\author[0000-0001-6580-6038]{Ayumu Shoshi}
\affiliation{Department of Earth and Planetary Sciences, Graduate School of Science, Kyushu University, 744 Motooka, Nishi-ku, Fukuoka 819-0395, Japan}
\email[show]{shoshi.ayumu.660@s.kyushu-u.ac.jp}

\author[0000-0002-8185-9882]{Masayuki Yamaguchi}
\affiliation{Department of Earth and Planetary Sciences, Faculty of Science, Kyushu University, 744 Motooka, Nishi-ku, Fukuoka 819-0395, Japan}
\affiliation{National Astronomical Observatory of Japan, 2-21-1 Osawa, Mitaka, Tokyo 181-8588, Japan}
\email{yamaguchi.masayuki.376@m.kyushu-u.ac.jp}  

\author[0000-0002-7951-1641]{Mitsuki Omura}
\affiliation{Department of Earth and Planetary Sciences, Graduate School of Science, Kyushu University, 744 Motooka, Nishi-ku, Fukuoka 819-0395, Japan}
\email{omura.mitsuki.362@s.kyushu-u.ac.jp}

\author[0000-0002-2062-1600]{Kazuki Tokuda}
\affiliation{Faculty of Education, Kagawa University, Saiwai-cho 1-1, Takamatsu, Kagawa 760-8522, Japan}
\email{tokuda.kazuki@kagawa-u.ac.jp}

\author[0009-0005-4458-2908]{Naofumi Fukaya}
\affiliation{Department of Physics, Nagoya University, Furo-cho, Chikusa-ku, Nagoya 464-8601, Japan}
\email{n.fukaya@a.phys.nagoya-u.ac.jp}

\author[0000-0002-1411-5410]{Kengo Tachihara}
\affiliation{Department of Physics, Nagoya University, Furo-cho, Chikusa-ku, Nagoya 464-8601, Japan}
\email{k.tachihara@a.phys.nagoya-u.ac.jp}

\author[0000-0002-0963-0872]{Masahiro N. Machida}
\affiliation{Department of Earth and Planetary Sciences, Faculty of Science, Kyushu University, 744 Motooka, Nishi-ku, Fukuoka 819-0395, Japan}
\email{machida.masahiro.018@m.kyushu-u.ac.jp}

\begin{abstract}
Recent ALMA observations with 0$\farcs$1 resolution reveal characteristic substructures in circumstellar disks around young Class I sources, providing clues to the early stages of morphological disk evolution.
In this paper, we applied \texttt{PRIISM} imaging to ALMA archival Band 6 continuum data of the circumstellar disk around the Class I protostar CrA~IRS~2, located in the Corona Australis molecular cloud, which is associated with an extended gas ring attributed to magnetic flux advection driven by interchange instability.
The dust continuum image with 1.5 times higher spatial resolution than conventional imaging revealed, for the first time,  the early-phase circumstellar disk with both inner central hole and outer ring-gap structures, making CrA~IRS~2 the youngest system exhibiting such features based on the bolometric temperature of $T_{\rm bol}$=235\,K. 
To examine planet-disk interaction as one possible origin of the outer ring-gap structure, we found the measured depth and width to be consistent with planet-disk interaction models, suggesting the existence of a giant planet with a mass of 0.1-1.8\,$M_{\rm Jup}$. 
The additional mechanism required for rapid planet formation could be the magnetic flux dissipation driven by the interchange instability, which suppresses MRI-driven turbulence and extends the dead zone, allowing efficient dust growth and the early formation of planets.
This system thus provides new insight into how substructures and planet formation can emerge during the early, accreting phase of disk evolution.
\end{abstract}

\keywords{Circumstellar disks (235); Protoplanetary disks(1300); Magnetic fields (994); Radio interferometry(1346)}

\section{Introduction}\label{sec:introduction}
Recent ALMA observations with $\lesssim$0$\farcs$1 resolution have revealed characteristic substructures, particularly ring-gaps, in circumstellar disks around young Class I sources \citep[e.g.,][]{Sheehan_2018,Shoshi_2024}.
To clarify the formation timescale of such substructures in circumstellar disks during the accretion phase, we applied high-resolution continuum imaging based on Sparse Modeling with \texttt{PRIISM} \citep[][]{Nakazato_2020,Nakazato_2020b} to 78 disks in the Ophiuchus molecular cloud \citep[][]{Shoshi_2025}.
Those images achieved spatial resolutions of 0$\farcs$02-0$\farcs$2 (a few au), and revealed that substructures predominantly appear in young stellar objects (YSOs) with bolometric temperatures $T_{\rm bol}>$200-300\,K. 
This range suggests that central stars have passed $\sim$0.2-0.4\,Myr after the star formed \citep[][]{Evans_2009}, and that substructures have also formed during this period.

This trend gives rise to a key outstanding question regarding the relationship between the substructures observed in young disks and the process of planet formation.
While planet-disk interaction has been proposed as one possible mechanism for the emergence of such substructures \citep[e.g.,][]{Dong_2015,Zhang_2018}, several hydrodynamical or chemical disk processes have also been suggested \citep[e.g.,][]{Youdin_2011,Takahashi_2016,Flock_2015}. 
Therefore, detailed studies of the morphology and the related to physical properties of substructures in young disks are crucial for elucidating their origin and evaluating links to planet formation.

In this paper, we focus on CrA~IRS~2 in the Corona Australis molecular cloud at the central J2000 coordinate ($19^{\rm h}01^{\rm m}41.48^{\rm s}$ $-36^\circ58^\prime31\farcs8$).
It is a Class I source with a spectral type K2 and a bolometric temperature of $T_{\rm bol}$=235\,K \citep[][]{Hsieh_2024}, located at a distance of 149.4\,pc \citep[][]{Galli_2020}. 
The circumstellar disk around the protostar is one of the targets observed in the CAMPOS project \citep[][]{Hsieh_2024} with a spatial resolution of 0$\farcs$13-0$\farcs$15 ($\sim$15\,au), which has revealed a dust ring-hole structure.
Moreover, recent ALMA observations of C$^{18}$O $J$=2--1 around the system at a spatial resolution of $\sim$200\,au have identified an extended gas ring with a diameter of $\sim$7000\,au, whose center was $\sim$5000\,au away from the protostar and the circumstellar disk \citep[][]{Tokuda_2023}.
The expansion of the gas ring can be attributed to magnetic flux advection driven by interchange instability, a form of magnetic buoyancy instability that has been theoretically predicted in protostellar environments \citep[e.g.,][]{Matsumoto_2017,Machida_2020}.
These findings motivate a detailed investigation of the circumstellar disk of CrA~IRS~2 to understand both the substructure formation during the accretion phase and the potential influence of interchange instability on disk evolution.

In this context, we reanalyze ALMA archival Band 6 continuum data of CrA~IRS~2 using super-resolution imaging with \texttt{PRIISM}.
We have achieved a higher spatial resolution image (0$\farcs$08), improved by a factor of 1.5 compared to the result obtained with conventional CLEAN imaging.
This paper is organized as follows.
Section~\ref{sec:obs_imaging} describes the data reduction and our imaging methods.
In Section~\ref{sec:result}, we present the high-resolution dust continuum image and estimate disk properties.
Section~\ref{sec:discussion} discusses the possible origin of the disk structures and the potential role of interchange instability.
Finally, Section~\ref{sec:summary} summarizes our conclusions.

\begin{figure*}[t]
    \centering
    \includegraphics[width=\linewidth]{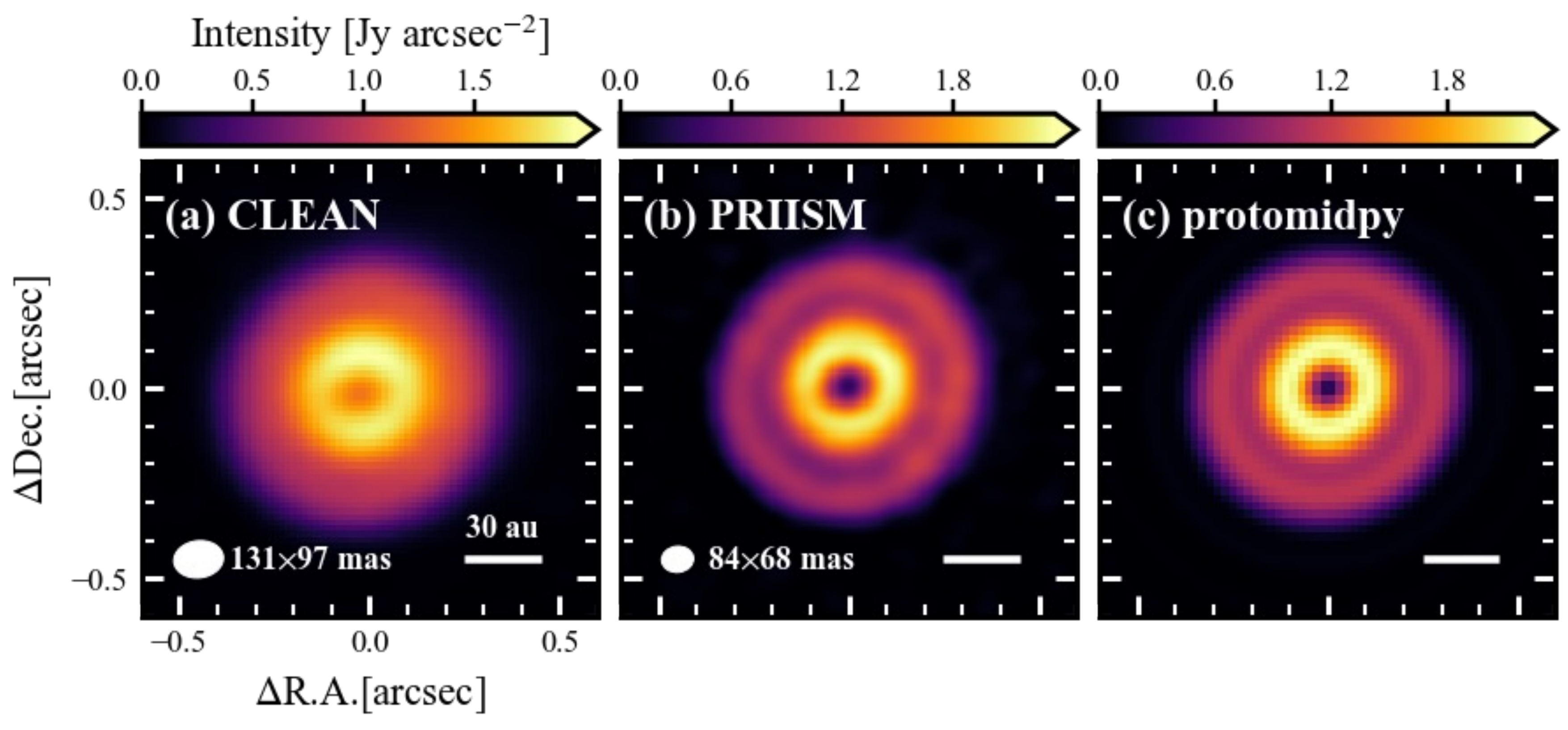}
    \caption{ALMA 1.3\,mm (Band 6) dust continuum around the Class I protostar CrA~IRS~2.
             The centr position corresponds to the J2000 coordinate ($19^{\rm h}01^{\rm m}41.48^{\rm s}$ $-36^\circ58^\prime31\farcs8$).
             (a) CLEAN image reconstructed with Briggs weighting (\texttt{robust}=0.5).
             The white ellipse denotes the synthesized beam.
             (b) PRIISM image reconstructed with hyper-parameters of $(\Lambda_l, \Lambda_{tsv})$=(10$^4$, 10$^{11}$).
             The white ellipse represents the effective spatial resolution $\theta_{\rm eff}$, determined by the point source injection method.
             (c) Disk model image generated using the best-fit geometric parameters and Gaussian kernel hyper-parameters with \texttt{protomidpy}.
              }
    \label{fig:continuum}
\end{figure*}

\section{ALMA Observation and Imaging}\label{sec:obs_imaging}
\subsection{Data Reduction}\label{subsec:reduction}
We analyzed ALMA Band 6 archival data of CrA~IRS~2 from Cycle 7 project 2019.1.01792.S (PI: Diego Mardones), part of the CAMPOS survey \citep[][]{Hsieh_2024}.  
The observations were conducted on 2021 July 10 with the C43-6/7 configuration, yielding an on-source integration time of 
$\sim$36\,seconds.
The raw data were calibrated using the ALMA pipeline in the Common Astronomy Software Applications package \citep[CASA;][]{CASA_2022}, version 6.1.1.  
Four spectral windows centered at 224, 226, 240, and 242\,GHz (each with 2\,GHz bandwidth) were employed for imaging the dust continuum.

\subsection{Imaging with CLEAN}\label{subsec:clean}
The continuum data were imaged using the \texttt{tclean} task in CASA version 6.1.0, with multi-frequency synthesis \citep[\texttt{nterm}=2;][]{Rau_2011}, the Cotton-Schwab algorithm \citep[][]{Schwab_1984}, and Briggs weighting (\texttt{robust}=0.5).
In order to improve image sensitivity, two rounds of self-calibration were applied by correcting gain errors.
Phase self-calibration (\texttt{calmode=p}) was first performed with an integration time equal to the on-source time (OST), followed by amplitude and phase calibration (\texttt{calmode=ap}) with an integration time of OST/5.
The achieved peak signal-to-noise ratio (SNR) on the final image is 320, which is 4.4 times higher than that of the image without the self-calibration.
The beam size is $0\farcs131\times0\farcs097$ with a position angle of -82.9$^\circ$, and the RMS noise level ($\sigma_{\rm dust}$) is 6.1\,mJy\,arcsec$^{-2}$ ($\sim$88\,$\mu$Jy\,beam$^{-1}$).

\subsection{Imaging with PRIISM}\label{subsec:priism}
We applied \texttt{PRIISM}\footnote{Python module for Radio Interferometry Imaging with Sparse Modeling; \url{https://github.com/tnakazato/priism}} imaging \citep[version 0.11.5;][]{Nakazato_2020,Nakazato_2020b} to the self-calibrated dataset processed with CASA 6.1.0, a technique that has been successfully applied to protoplanetary disks in previous studies \citep[e.g.,][]{Yamaguchi_2025,Shoshi_2025b}.
This technique employs $\ell_1$+TSV regularized imaging using a 10-fold cross-validation (CV) scheme \citep[for details, see][]{Yamaguchi_2024,Shoshi_2025}. 
The cost function is minimized, comprising a chi-squared error term and two regularization terms, the $\ell_1$-norm and the total squared variation (TSV).
The chi-squared term quantifies the difference between the observed visibilities and those predicted from the model image via Fourier transformation.
The $\ell_1$-norm controls the sparsity of the brightness distribution and its total flux, whereas TSV controls smoothness by changing pixel-to-pixel variations.
The hyper-parameters $\Lambda_l$ and $\Lambda_{tsv}$ associated with the $\ell_1$-norm and TSV control the relative contributions of the two regularization terms with respect to the observational data.

We optimized these hyper-parameters over a wide range using the 10-fold CV approach and initially selected the combination $(\Lambda_l, \Lambda_{tsv})=(10^4, 10^{10})$ as the best parameter set.
However, the resulting image showed artificial clumpy structures.
We therefore conservatively increased $\Lambda_{tsv}$ by one order of magnitude to $(\Lambda_l, \Lambda_{tsv})=(10^4, 10^{11})$, which produced a more stable image with a smoothed brightness distribution (for details, see Appendix~\ref{sec:tsv_setting}).

Unlike the conventional CLEAN algorithm, the PRIISM imaging process does not undergo beam convolution. 
Therefore, to measure its effective spatial resolution $\theta_{\rm eff}$, we applied the point-source injection method \citep[][]{Yamaguchi_2021}.
An artificial point source with 5\% of the total flux was injected into the observed visibility, whose position was centered in an emission-free region near the central star within the maximum recoverable scale ($\sim$2.0\,arcsec). 
The reconstructed point source is approximately represented by a Gaussian function on the image domain. 
Its full width at half maximum (FWHM) is set to the effective spatial resolution $\theta_{\rm eff}$.
The resulting resolution was $0\farcs084 \times 0\farcs068$ at a position angle of $96.6^\circ$, achieving about 1.5 times better improvement to that of the CLEAN image.

\section{Results}\label{sec:result}
\subsection{Multiple Ring Structures}\label{subsec:structure}
Figure~\ref{fig:continuum} shows the 1.3\,mm dust continuum images.  
A ring-hole structure in the inner region is already shown in the CLEAN image (Figure~\ref{fig:continuum}a). 
In contrast, the 1.5 times higher spatial resolution achieved through the \texttt{PRIISM} imaging reveals an additional substructure of the ring-gap in the image (Figure~\ref{fig:continuum}b).
Thus, the CrA~IRS~2 disk is found to harbor the two prominent substructures with inner ring-hole and outer ring-gap structures.
The previous disk imaging survey in \citet{Shoshi_2025} has identified the disk substructures emerging from $T_{\rm bol}\gtrsim$200-300\,K.
Among them, the CrA~IRS~2 is the youngest one with multiple ring structures in terms of its low bolometric temperature of $T_{\rm bol}$=235\,K, compared to the other Class I disks with similar structures \citep[e.g.,][]{Sheehan_2018,Segura-Cox_2020}.

To further confirm the image fidelity of the disk substructures, we performed the one-dimensional non-parametric visibility fitting approach with \texttt{protomidpy}\footnote{\url{https://github.com/2ndmk2/protomidpy}} \citep[][]{Aizawa_2024} and produced an axisymmetric disk model. 
Using Markov Chain Monte Carlo (MCMC) sampling, we derived best-fit values for the disk geometry parameters and the two hyper-parameters of the Gaussian Process kernel \citep[for details, see][]{Aizawa_2024}. 
The results are central offsets $(\Delta x_{\rm cen}, \Delta y_{\rm cen})=(-9.28^{+0.07}_{-0.07}, 2.38^{+0.06}_{-0.07})$ in units of mas, inclination angle $\cos i=0.946^{+0.001}_{-0.001}$, position angle ${\rm PA}={131.61^{+0.30}_{-0.27}}^\circ$, and two hyper-parameters $\log_{10}\alpha=-1.28^{+0.07}_{-0.06}$ and $\gamma=0.071^{+0.002}_{-0.002}$.
The value of $\gamma$ defines the effective spatial scale of the reconstructed brightness profile. 
The measurement of PA is consistent with the value in \citet{Hsieh_2024} within the error.
The model image shown in Figure~\ref{fig:continuum}(c) clearly reproduces the central hole and the ring gap structure behind it, which is also confirmed in the PRIISM image. 
The agreement between the two independent methods supports the robustness of the detected substructures.

Both direct approaches, using \texttt{PRIISM} and \texttt{protomidpy}, to the observed visibility show a similar brightness distribution.
However, we should consider the differences in the assumptions and the treatment of the observed visibility between the two imaging methods.
The \texttt{protomidpy} approach generates an axisymmetric disk structure by compressing the observed visibilities into one dimension and fitting them with a Fourier-Bessel series. 
This procedure inevitably averages out any azimuthal structures present in the data \citep[for details, see][]{Aizawa_2024}.
In contrast, the PRIISM imaging reconstructs the two-dimensional structure directly from the original two-dimensional visibilities (see \S\ref{subsec:priism}). 
This allows us to account for local brightness variations and asymmetries and to obtain a more model-independent and high-fidelity evaluation of the gap width and depth, even after azimuthal averaging. 
The advantage is particularly more effective in the case of CrA~IRS~2, which exhibits a small asymmetric brightness distribution on the southwestern side of the outer ring (see Appendix~\ref{sec:tsv_setting}). 
In addition, a widely used metric exists to measure its effective spatial resolution, and previous studies have validated that \texttt{PRIISM} can produce high-fidelity images as demonstrated in \citet{Yamaguchi_2021,Yamaguchi_2024} and \citet{Shoshi_2025}.
Therefore, in the following sections, we adopt the PRIISM image rather than the \texttt{protomidpy} approach for characterizing the substructures. 

\begin{figure}[t]
    \centering
    \includegraphics[width=\linewidth]{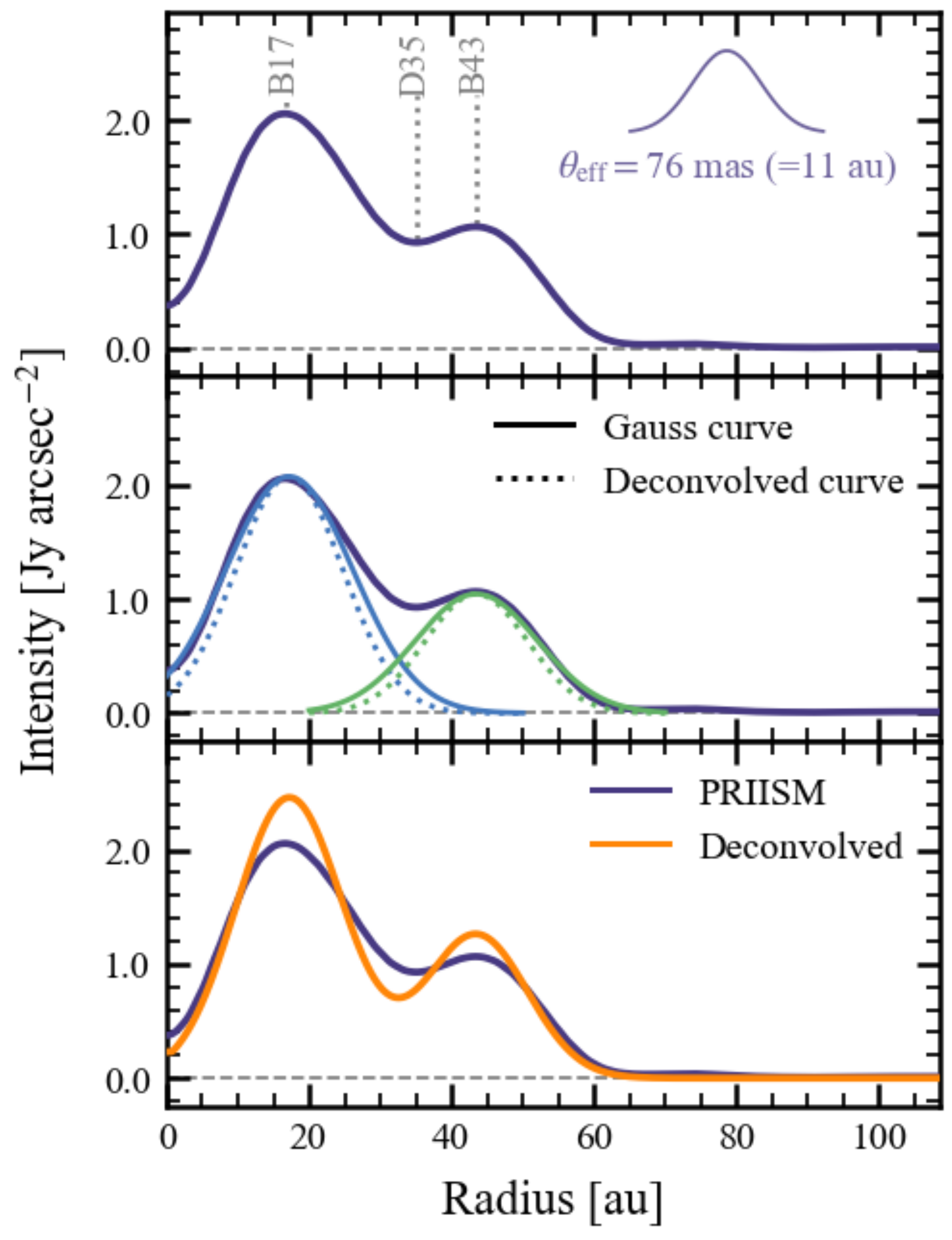}
    \caption{Radial intensity profile averaged over the full azimuthal angle.
             The violet solid curves in all panels are the profile of the PRIISM image.
             In the middle panel, the blue and green solid curves represent the two components determined by a multiple-Gaussian fitting approach (see \S\ref{subsec:morphology}).
             The corresponding deconvolved profiles, with the deconvolved standard deviations $\sigma_{\rm dec}$ and the original amplitudes $I_0$ described in Table~\ref{table:ring}, are shown as dotted lines.
             The deconvolved profile corresponds to the orange curve in the bottom panel (see \S\ref{subsec:gap_formation}).
             All the radial profiles are linearly interpolated onto radial grid points spaced by 0.1\,au using \texttt{interpolate.interpld} from the \texttt{SciPy} module.
             }
    \label{fig:profile}
\end{figure}

\subsection{Disk Morphology}\label{subsec:morphology}
We quantify the disk geometry and characterize the substructures detected in the PRIISM image.
We first determined the peak positions along the outer ring and then fitted them with an open ellipse \citep[][]{Yamaguchi_2021, Shoshi_2024}, obtaining an inclination angle ($i_{\rm disk}$) of $21.2\pm0.3^\circ$ and a position angle (PA) of $145.3\pm1.0^\circ$.
Note that the difference from the results derived by \texttt{protomidpy} in \S\ref{subsec:structure} arises from the use of different imaging methods.
Figure~\ref{fig:continuum}(c) was derived from one-dimensional visibility fitting based on a disk model represented by a Fourier-Bessel series \citep[for details, see][]{Aizawa_2024}, whereas the \texttt{PRIISM} image in Figure~\ref{fig:continuum}(b) was directly reconstructed from the visibilities without assuming any disk model, indicating that the imaging process is model-independent \citep[e.g.,][]{Yamaguchi_2021, Yamaguchi_2024}.
Therefore, the discrepancy in the inclination angle ($\sim$3\% of the entire inclination angle of 90$^\circ$) is attributed to methodological differences.
We have verified that adopting either set of orientation parameters for the deprojection changes the brightness distributions by at most $\sim 2.7\%$ (i.e., $\sim 0.06\ {\rm Jy\ arcsec^{-2}}$) of the peak intensity in the PRIISM image, and thus this discrepancy is not critical for the subsequent analysis.
In this paper, we conservatively adopt $i_{\rm disk}=21.2\pm0.3^\circ$ and PA=$145.3\pm1.0^\circ$ measured from the model-independent PRIISM image to construct the azimuthally averaged radial intensity profile shown in Figure~\ref{fig:profile}.

We then characterized the disk using the curve-growth method \citep[e.g.,][]{Yamaguchi_2024} to derive the total flux density $F_\nu$ and disk radius $R_{95\%}$, which defines the radius enclosing 95\% of the flux.
From the PRIISM image, we obtained $F_\nu = 513.9 \pm 51.3$\,mJy and $R_{95\%} = 57.8 \pm 5.3$\,au, which is relatively larger than those of other Class I and II disks in the Ophiuchus and Taurus regions in \citet{Shoshi_2025} and \citet{Yamaguchi_2024}.
The uncertainties include a 10\% absolute flux calibration error for total flux and the standard deviation of $\theta_{\rm eff}$ for the disk radius.
The total flux of the PRIISM image is comparable to that of the CLEAN image in the region with dust emissions larger than 5$\sigma_{\rm dust}$.

\begin{table}[t]
    \centering
    \caption{Results of a multiple Gaussian fitting}
    \begin{tabular}{lccccc}
    \hline
    Label & \multicolumn{3}{c}{(Original)} & \multicolumn{2}{c}{(Deconvolved)} \\
          & $r_0$ & $I_0$           & $\sigma$ & $I^{\rm dec}_0$   & $\sigma_{\rm dec}$ \\
          & $\left[{\rm au}\right]$ & $\left[{\rm Jy\,arcsec^{-2}}\right]$ & $\left[{\rm au}\right]$ & $\left[{\rm Jy\,arcsec^{-2}}\right]$ & $\left[{\rm au}\right]$ \\
    \hline
    \hline
    B17 & 17.2 & 2.1 & 8.9 & 2.5 & 7.5 \\
    B43 & 43.5 & 1.1 & 8.6 & 1.3 & 7.2 \\
    \hline
    \end{tabular}
    \label{table:ring}
\end{table}

\subsection{Substructure Characterization}\label{subsec:subcha}
The radial intensity profile shown in Figure~\ref{fig:profile} exhibits two peaks corresponding to the inner and outer ring structures.
To characterize this profile, we modeled it as a sum of multiple Gaussian components.
Each component follows the form $I_\nu^{\rm gauss}(r)=I_0\exp\left[-(r-r_0)^2/2\sigma^2\right]$, where $I_0$ is the peak intensity, $r_0$ is the peak position, and $\sigma$ denotes the standard deviation.
We obtained the parameters using the MCMC approach with \texttt{emcee} \citep{Foreman-Mackey_2013}, which employed flat priors and ran 100 walkers for 2000 steps (with the initial 500 steps discarded as burn-in).
Table~\ref{table:ring} summarizes the best-fit parameters of the Gaussian components for the inner and outer rings, and the middle panel of Figure~\ref{fig:profile} displays the corresponding profiles. 
Following \citet{ALMA_2015}, the radial locations of the inner and outer rings are labeled as B17 and B43, respectively, and the location of the local minimum in the profile ($r$=35\,au) is defined as D35.
The multiple Gaussian fitting with MCMC yields typical uncertainties of at most $\sim 0.01$\,au in $r_0$ and $\sigma$, and $\sim 10^{-3}\ {\rm Jy\ arcsec^{-2}}$ in $I_0$.
We have confirmed that, even when these uncertainties are propagated to the derived gap properties discussed below, the resulting errors remain much smaller than the values themselves and are therefore not included in the subsequent analysis.

We then measured the gap depth $\delta_{\rm I}$ and normalized gap width $\Delta_{\rm I}$ for the gap D35, following Case 1 of \citet{Yamaguchi_2024}, which is based on the definition used for planet-disk interaction simulations \citep[][]{Zhang_2018}.
Note that the central hole could be related to the interactions with planets. 
Still, its structure remains unresolved due to the constraints of sensitivity and spatial resolution, and we do not discuss it further in this paper.
The gap depth is defined as the intensity ratio between the ring and the gap, $\delta_{\rm I}=I_\nu(r_{\rm ring})/I_\nu(r_{\rm gap})$,
where $r_{\rm ring}$ and $r_{\rm gap}$ are the locations of B43 and D35, respectively, and $I_\nu(r)$ is the intensity at each location.
The normalized gap width is given by $\Delta_{\rm I}=(r_{\rm out}-r_{\rm in})/r_{\rm out}$, where $r_{\rm in}$ and $r_{\rm out}$ are the radial positions at which $I_\nu(r)=I_{\rm edge}\equiv0.5[I_\nu(r_{\rm ring})+I_\nu(r_{\rm gap})]$.
Applying this formula, we obtained $\delta_{\rm I} = 1.15$ and $\Delta_{\rm I} = 0.19$.

To mitigate the effect of the smoothness determined by the effective spatial resolution $\theta_{\rm eff}$, assuming that the intrinsic internsity profile is composed of Gaussian functions, we applied the deconvolution of the PRIISM image.
Following \citet{Dullmond_2018} and using the results of a multiple-Gaussian fitting, each deconvolved component is expressed as
$I_\nu^{\rm gauss, dec}(r) = I_0^{\rm dec} \exp\left[-(r - r_0)^2 / 2\sigma_{\rm dec}^2\right]$,
where $\sigma_{\rm dec} = \sqrt{\sigma^2 - \sigma_{\rm eff}^2}$ is the deconvolved standard deviation, and $I_0^{\rm dec} = \sigma I_0 / \sigma_{\rm dec}$ is the peak intensity adjusted to conserve total flux.
Table~\ref{table:ring} lists the deconvolved parameters for B17 and B43, and the red curve in the bottom panel of Figure~\ref{fig:profile} shows the resulting deconvolved profile.
Using this deconvolved profile, we estimated a gap depth of $\delta_{\rm I, dec}$=1.79 and a normalized width of $\Delta_{\rm I, dec}$=0.26 in the same way.

To the best of our knowledge, although the basic idea is conceptually similar to the deconvolution approach of \citet{Dullmond_2018}, this is the first study in which the intrinsic gap properties have been inferred from the dust continuum image produced without the conventional beam convolution process.
Note that \citet{Dullmond_2018} described only the deconvolved ring width and amplitude. 
Based on the gap widths and depths measured from both the original and deconvolved radial profiles, we discuss the possibility of planet-disk interaction in \S\ref{subsec:gap_formation}.

\section{Discussion}\label{sec:discussion}
\subsection{The Youngest Class I Disk with Multiple Ring Structures}\label{subsec:youngest}
We report the first identification of the outer ring-gap structure, resulting in the circumstellar disk with two distinctive ring structures around the Class I protostar CrA~IRS~2.
Recent ALMA observations at high spatial resolution have revealed ring structures in several Class 0/I disks \citep[e.g.,][]{Maureira_2024,Sheehan_2020,Sai_2020}. 
In most cases, these disks exhibit a single ring morphology \citep[e.g.,][]{Sheehan_2017,Maureira_2024}, and those with multiple rings account for less than half of the sample. 
In addition, nearly edge-on disks may be misclassified with respect to their evolutionary stage \citep[][]{Furlan_2016,Shoshi_2025}. 
When these effects are taken into account, the nearly face-on Class I disks exhibiting multiple ring structures are limited to three systems: CrA~IRS~2, ISO-Oph~54, and HL~Tau \citep[][]{Sheehan_2018,ALMA_2015}. 
Since substructures are ubiquitous in Class II disks \citep[e.g.,][]{Andrews_2018,Cieza_2021}, these three systems provide crucial examples for studying the origin of multiple rings during the accretion phase. 
In this subsection, we compare these systems to investigate the formation processes of multiple rings. 
For Oph~IRS~63 ($T_{\rm bol}=348$\,K), \citet{Segura-Cox_2020} reported a Class I disk with multiple rings by subtracting a disk model from the dust continuum image, but \citet{Flores_2023} also noted that there is at least a single diffuse annular feature, rather than clearly resolved multiple rings.
In contrast, for CrA~IRS~2, ISO-Oph~54, and HL~Tau, the ring and gap structures are relatively clearly visible in both the continuum images and the radial intensity profiles, allowing a more straightforward and consistent comparison.
For these reasons, we conservatively did not include Oph~IRS~63 ($T_{\rm bol}=348$\,K) as a candidate with multiple rings for comparison with other Class I disks.

Compared to ISO-Oph~54 and HL~Tau, the most notable characteristic of CrA~IRS~2 is its youth. 
The bolometric temperature of CrA~IRS~2 is 235\,K \citep[][]{Hsieh_2024}, corresponding to the middle Class I stage, which typically occurs at $\sim$0.2-0.4\,Myr after protostar formation \citep[][]{Evans_2009}. 
In contrast, the bolometric temperatures of ISO-Oph~54 and HL~Tau are 380\,K and 576\,K, respectively, indicating that they are in more evolved evolutionary phases \citep[][]{Dunham_2015,Chen_1995}. 
Therefore, CrA~IRS~2 could be the youngest disk exhibiting multiple ring structures, which were likely formed rapidly during the accretion phase.

Considering their relative ages, we evaluate the transition of disk morphology among the three systems.
First, the dust disk radius of CrA~IRS~2 ($\sim$57.8\,au) is less than half that of ISO-Oph~54 \citep[$\sim$119\,au;][]{Cieza_2021} and HL~Tau \citep[$\sim$140\,au;][]{ALMA_2015}, suggesting that the more evolved disk is larger among the three.
Note that these systems reside in different star-forming regions, and the variation in disk size may be influenced by environmental factors \citep[e.g.,][]{Hendler_2020}.

Furthermore, CrA~IRS~2 exhibits two ring structures separated by approximately 26\,au, whereas ISO-Oph~54 shows four rings with an average spacing of about 22\,au \citep[][]{Cieza_2021}, and HL~Tau shows seven rings with a mean spacing of 13\,au \citep[][]{ALMA_2015}.
Among these three systems, the more evolved disks (ISO-Oph~54 and HL~Tau) tend to host more rings, possibly scaling with the overall disk size rather than with evolution alone.
In addition, the limited sample size prevents us from establishing any robust trend.
Moreover, the rings in HL~Tau are unevenly spaced, with separations of $\sim$18\,au in the inner region within $\sim$40\,au and $\sim$10\,au or smaller in the outer region.
These comparisons suggest that CrA~IRS~2 may correspond to an earlier evolutionary stage of systems such as ISO-Oph~54 and HL~Tau, but further investigation of its formation mechanisms will require examining the gas distribution and kinematics.

\subsection{Qualitative Evaluation of Substructure Formation Mechanisms}\label{subsec:subst_origin}
As described in \S\ref{subsec:youngest}, the inner ring-hole and the outer ring-gap structures could be formed rapidly in the relatively early Class I phase.
In the discussion on the mechanism of the rapid substructure formation of CrA~IRS~2, we should consider the existence of the interchange instability.
\citet{Tokuda_2023} revealed an extended bubble of C$^{18}$O gas, which was likely to be driven by magnetic advection associated with the interchange instability and connected to the circumstellar disk with two ring structures through SO gas emission.
The non-ideal MHD simulation in \citet{Machida_2025} expected the magnetic field within the disk to be below $\sim$10\,mG (plasma $\beta\gtrsim10^4$) in $3\times10^4$ years after protostar formation.
These considerations underscore the need for formation mechanisms that operate effectively in weak magnetic fields.

Several mechanisms have been proposed to explain ring-gap structures, such as secular gravitational instability \citep[][]{Takahashi_2014}, MHD wind \citep[e.g.,][]{Suriano_2019}, snow lines of molecules or dust sintering \citep[e.g.,][]{Okuzumi_2016}, and so on.
However, these models are primarily developed for Class II disks and do not include external mass infall onto the disk.
In particular, the MHD wind model by \citet{Suriano_2019} produces prominent multiple ring-gap structures only under relatively strong magnetization (plasma $\beta\sim10^3$), whereas the magnetic field in CrA~IRS~2 is expected to be much weaker \citep[$\beta\gtrsim10^4$, see Figures~5 and 7 in][]{Machida_2025}, making this scenario difficult to apply directly to our target.
For rapid substructure formation during the accretion phase, plausible mechanisms include binary companions, disk winds with external mass infall, dust growth fronts, and embedded planets.

Binary companions can generate asymmetric or distorted ring structures through gravitational torques on the circumbinary disk \citep[e.g.,][]{Takakuwa_2017,Matsumoto_2019,Saiki_2020}.
The PRIISM image shows no large-scale asymmetries, and the overall axisymmetric morphology makes a binary-induced origin unlikely, even if a very close companion cannot be entirely excluded.

The model of \citet{Takahashi_2018} includes external infall and MRI-driven disk winds that remove gas from the inner regions and push dust outward, producing only a single ring-gap structure that forms inside-out. 
This behavior is inconsistent with CrA~IRS~2, which exhibits a clear inner ring in addition to the outer one. 
In addition, although the angular resolution of the C$^{18}$O and SO data in \citet{Tokuda_2023} does not resolve the immediate vicinity of the inner ring, the detection of extended C$^{18}$O and SO emission around the system indicates that the molecular gas may exist in the inner region.
This makes it less likely that the inner region contains less gas, as expected in the model of \citet{Takahashi_2018}.
Still, higher-angular-resolution molecular-line observations are required to assess the gas content near the inner ring directly.

\citet{Ohahsi_2021} showed that inside-out dust coagulation in young Class 0/I disks can produce a bright ring corresponding to a dust growth front. 
However, their model predicts rapid inward dust drift within the growth front, creating an inner cavity rather than additional rings, which is inconsistent with the two-ring morphology of CrA~IRS~2.

Overall, the scenarios involving binary companions, disk winds with external infall, and dust growth fronts rely only on qualitative comparisons with the brightness distribution in the PRIISM image and therefore provide limited constraints. 
To discriminate more robustly among these mechanisms and quantify their relative importance, future high-sensitivity and high-resolution molecular line observations of CrA~IRS~2 would be essential for better constraining the gas distribution, kinematics, and turbulence level in the disk.
Therefore, in \S\ref{subsec:gap_formation}, based on the quantitative measurements derived in \S\ref{subsec:subcha}, we mainly verify whether planet-disk interaction can account for the detected ring structures.

\subsection{Possible Formation of Ring-Gap Structure Induced by Planet-disk Interaction}\label{subsec:gap_formation}
\subsubsection{Comparison with Planet-disk Interaction Models}\label{subsubsec:comparison}
\begin{figure}[t]
    \centering
    \includegraphics[width=\linewidth]{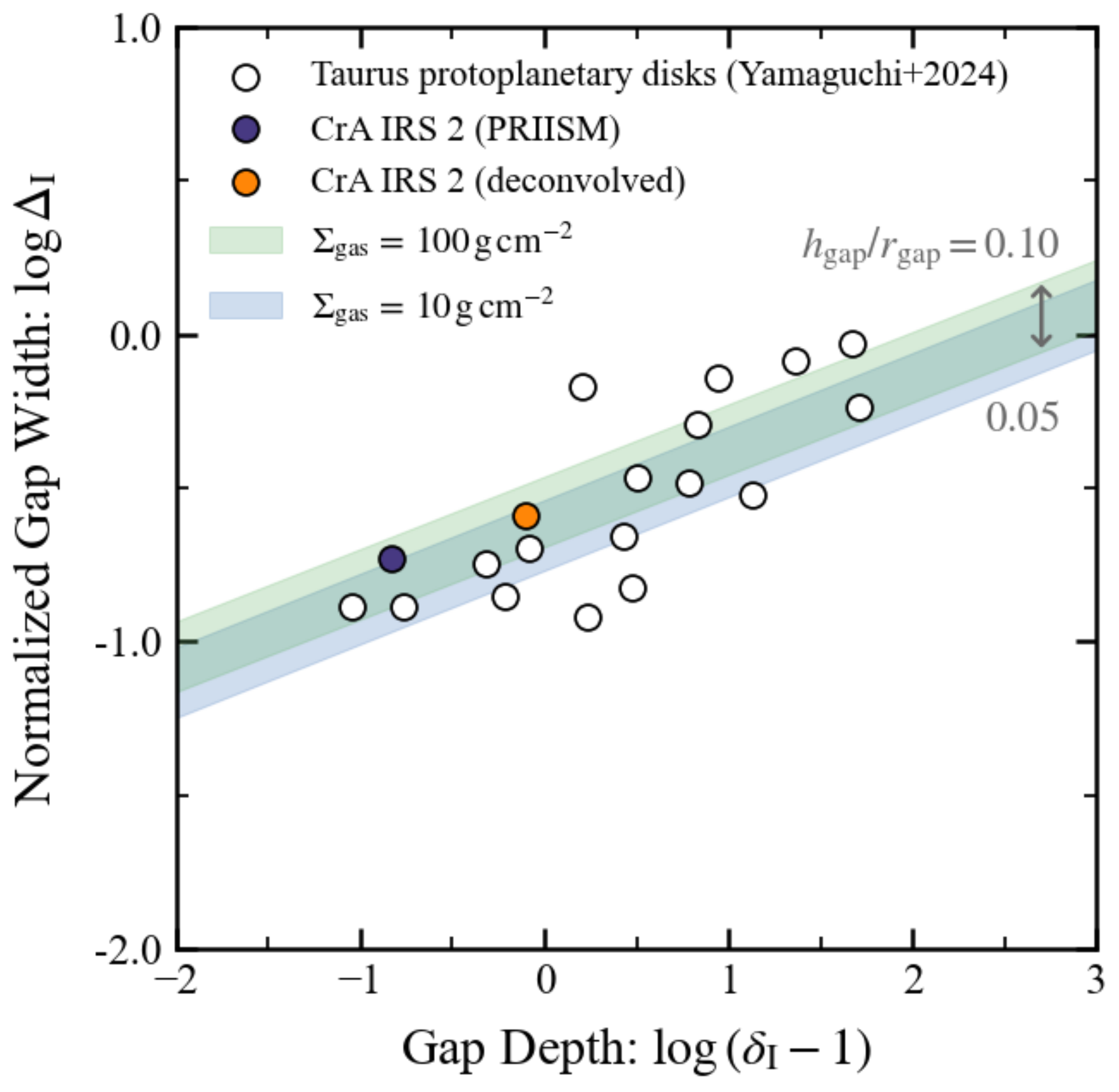}
    \caption{Relationship between gap width and depth based on the model from \citet{Zhang_2018} (shaded region), overlaid with the gap properties of the circumstellar disk around CrA~IRS~2 (this study) and of Taurus protoplanetary disks reported by \citet{Yamaguchi_2024}.
             Using fixed values of a maximum dust grain size of $s_{\rm max}$=0.1\,mm and a viscosity parameter of $\alpha_{\rm vis}=10^{-3}$, we present models for gas surface densities $\Sigma_{\rm gas}$ of 100\,g\,cm$^{-2}$ (green) and 10\,g\,cm$^{-2}$ (blue), with the disk scale height ranging from 0.05 to 0.10, which is used in \citet{Zhang_2018}.
             }
    \label{fig:scatter}
\end{figure}

We compare the observed gap in CrA~IRS~2 with the predictions of the planet-disk interaction model by \citet{Zhang_2018}, who conducted two-dimensional hydrodynamic simulations including both gas and dust components. 
They derived an empirical relation between the gap depth $\delta_{\rm I}$ and normalized width $\Delta_{\rm I}$ as
\begin{align}
    \Delta_{\rm I}=A\left[0.635\left(\frac{h_{\rm gap}}{r_{\rm gap}}\right)^{2.63}\left(\frac{\alpha_{\rm vis}}{10^{-3}}\right)^{0.07}\left(\frac{\delta_{\rm I}-1}{C}\right)^{1/D}\right]^B,\label{eq:width_depth}
\end{align}
where $h_{\rm gap}/r_{\rm gap}$ is the disk aspect ratio and $\alpha_{\rm vis}$ is the viscosity at the gap location. 
The constants $A$, $B$, $C$, and $D$ depend on the gas surface density $\Sigma_{\rm gas}$ and the maximum dust grain size $s_{\rm max}$.
We adopt fixed values of $A=1.11,B=0.29, C=0.0478,$ and $D=1.23$ in the case of $s_{\rm max}=0.1$\,mm, which was based on ALMA polarization observations toward other disks by \citet{Bacciotti_2018}, and $\alpha_{\rm vis} = 10^{-3}$ \citep[DSD1 model;][]{Zhang_2018}, according to \citet{Yamaguchi_2024}.
In Figure~\ref{fig:scatter}, the shaded regions represent the model predictions for $\Sigma_{\rm gas} $=10 and 100\,g\,cm$^{-2}$, and $h_{\rm gap}/r_{\rm gap}$ set from 0.05 to 0.10, corresponding to the adopted range of disk aspect ratios used in the DSD1 models of \citet{Zhang_2018}.

We compared the models in Figure~\ref{fig:scatter} with the widths and depths of the gap structure D35 in the original and decovolved cases estimated in \S\ref{subsec:subcha}, finding that both cases are consistent with the prediction of \citet{Zhang_2018}.
This implies that the gap structure D35 in CrA~IRS~2 can be attributed to interactions with an embedded planet, suggesting that a young planet may have already formed during the Class I stage.

\begin{figure}[t]
    \centering
    \includegraphics[width=\linewidth]{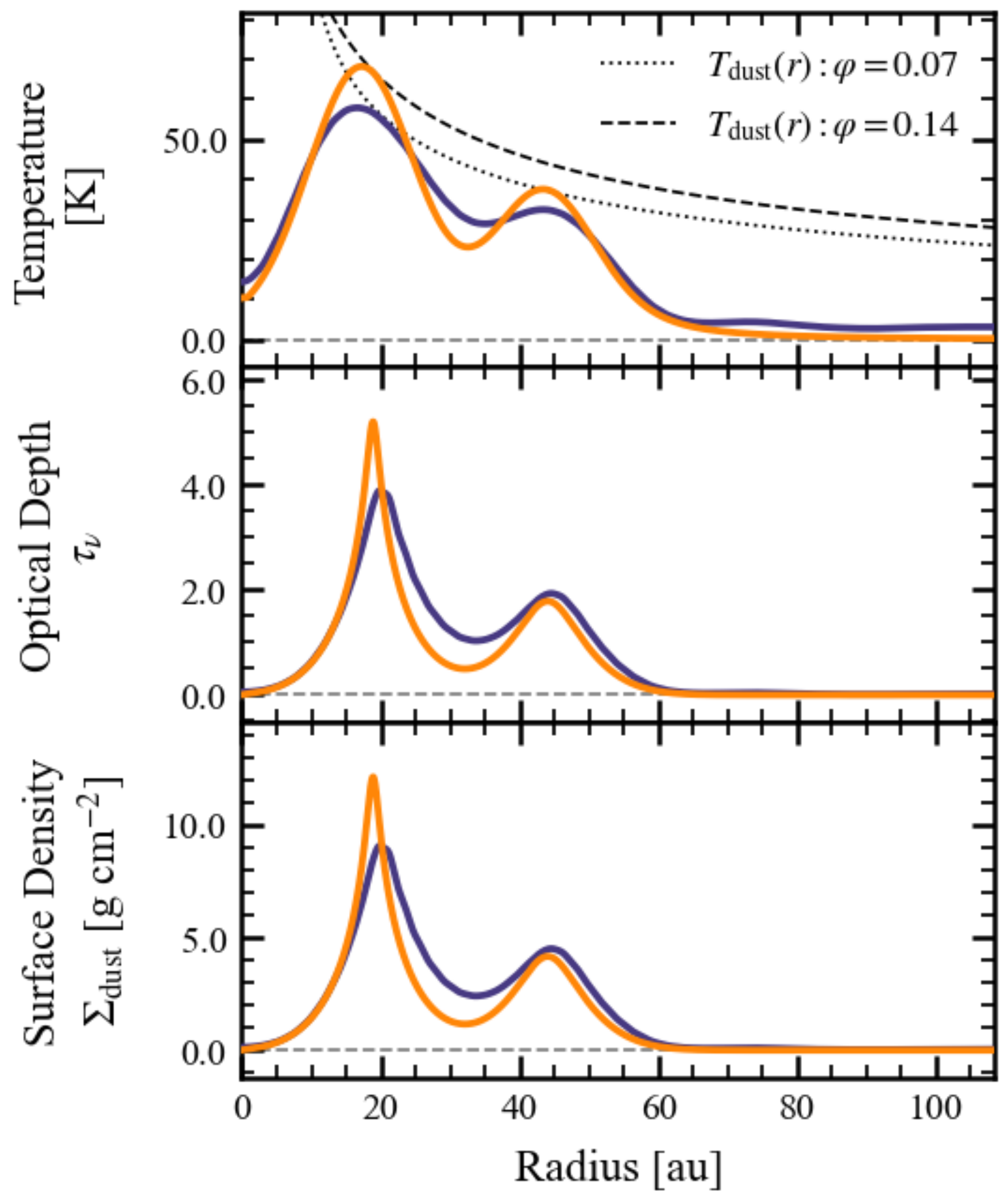}
    \caption{(Top) Radial profiles of the brightness temperature in the original (PRIISM; purple) and the deconvolved (orange) cases.
             The gray dotted and dashed lines represent disk temperature models $T_{\rm dust}(r)$ of $\varphi$=0.07 and 0.14, respectively.
             (Middle) Optical depths of two cases $\tau_\nu$. 
             The purple line was derived from the original brightness temperature profile and the dust disk temperature model of $\varphi$=0.07. 
             In contrast, the orange line was estimated from the deconvolved profile and the model of $\varphi$=0.14.
             (Bottom) Dust surface densities of the two models $\Sigma_{\rm dust}(r)$ estimated with the dust opacity $\kappa_\nu$=0.43\,cm$^2$\,g$^{-1}$ \citep[][]{Birnstiel_2018}.}
    \label{fig:placeholder}
\end{figure}

\subsubsection{Planet Mass Estimation from Gap Properties}\label{subsubsec:planet_mass}
We estimated the planet mass $M_{\rm p}$ responsible for the observed gap using the empirical relation derived by \citet{Zhang_2018} as
\begin{align}
\frac{M_{\rm p}}{M_\ast}=0.115\left(\frac{\Delta_{\rm I}}{A}\right)^{1/B}\left(\frac{h_{\rm gap}}{r_{\rm gap}}\right)^{0.18}\left(\frac{\alpha_{\rm vis}}{10^{-3}}\right)^{0.31},\label{eq:planet_mass}
\end{align}
where $M_\ast$ is the stellar mass, and $A$ and $B$ are the same constants in Equation~(\ref{eq:width_depth}). 
We adopted $s_{\rm max}$=0.1\,mm \citep[][]{Bacciotti_2018} and $\alpha_{\rm vis}=10^{-3}$, consistent with the DSD1 model of \citet{Zhang_2018}.

The constants $A$ and $B$ in Equation~(\ref{eq:planet_mass}) depend on the model of $\Sigma_{\rm gas}$.
To constrain $\Sigma_{\rm gas}$, we first derived the dust surface density $\Sigma_{\rm dust}(r)$ using the radiative transfer equation, $\Sigma_{\rm dust}(r)=\tau_\nu/\kappa_\nu=-\ln[1-I_\nu(r)/B_\nu (T_{\rm dust}(r))]/\kappa_\nu$, with $\kappa_\nu = 0.43$\,cm$^2$\,g$^{-1}$ \citep[DSHARP opacity model;][]{Birnstiel_2018}.
The top panel of Figure~\ref{fig:placeholder} shows the brightness temperatures for the original and deconvolved cases.
The dust temperature $T_{\rm dust}(r)$ was assumed as $T_{\rm dust}(r)=(\varphi L_\ast/8\pi r^2\sigma_{\rm SB})^{1/4}$ \citep[e.g.,][]{Chiang_1997}, where $L_\ast$ is the stellar luminosity of $L_\ast=4.3\,L_\odot$ \citep[][]{Fiorellino_2023}.
The flaring angle, $\varphi$, was increased from 0.05 in increments of 0.01, and the minimum value for which the condition $I_\nu (r) < B_\nu \left[T_{\rm dust}(r)\right]$ is satisfied at all radii was adopted in order to prevent the optical depth $\tau_\nu=-\ln[1-I_\nu(r)/B_\nu (T_{\rm dust}(r))]$ from diverging to infinity.
As a result, $\varphi = 0.07$ was chosen for the original profile and $\varphi = 0.14$ for the deconvolved profile, whose dust temperatures were represented as the black dotted and dashed lines in the top panel of Figure~\ref{fig:placeholder}.
The middle panel shows the curves of the optical depths $\tau_\nu(r)$ without divergence to infinity, indicating that suitable values of the flaring angles $\varphi$ could be determined.
As shown in the bottom panel of Figure~\ref{fig:placeholder}, the mean dust surface densities from 1.1\,$r_{\rm gap}$ to 2.0\,$r_{\rm gap}$ are $\Sigma_{\rm dust}\sim$1.9\,g\,cm$^{-2}$ in the original case and $\Sigma_{\rm dust}\sim$1.7\,g\,cm$^{-2}$ in the deconvolved case, corresponding to the DSD1 model of the gas surface density $\Sigma_{\rm gas} \sim 100$\,g\,cm$^{-2}$ based on Figure~18 of \citet{Zhang_2018}.
The mean value of the dust surface density should be regarded as an upper limit since we determined the lower limit of the dust temperature $T_{\rm dust}(r)$ based on the minimum value of the flaring angle.
However, the change in the surface density would only weakly modify the coefficients $A$ and $B$ in Equation~(\ref{eq:planet_mass}) and would not substantially affect the estimation of the planet mass.

We substituted $\Delta_{\rm I}$ (or $\Delta_{\rm I, dec}$) estimated in \S\ref{subsec:gap_formation}, $h_{\rm gap}/r_{\rm gap}=0.07$ for the original profile and $h_{\rm gap}/r_{\rm gap}=0.10$ for the deconvolved profile, $\alpha_{\rm vis}=10^{-3}$, the constants $A$ and $B$ for $\Sigma_{\rm gas}=100\,{\rm g\,cm^{-2}}$, and the stellar mass $M_\ast=1.4\pm0.3\,M_\odot$ \citep[][]{Nisini_2005,Fiorellino_2023} into Equation~(\ref{eq:planet_mass}).
Note that $h_{\rm gap}/r_{\rm gap}=0.10$ corresponds to the upper end of the aspect–ratio range explored by the DSD1 models of \citet{Zhang_2018}, and even extrapolating to $h_{\rm gap}/r_{\rm gap}=0.14$ would increase the inferred planet mass by only $\sim 6\%$.
We finally obtained a planet mass of $M_{\rm p}=0.22^{+0.33}_{-0.14}\,M_{\rm Jup}$ from the PRIISM image ($M_{\rm p, dec}=0.71^{+1.05}_{-0.44}\,M_{\rm Jup}$ in the deconvolved case), which is considered to cause the gap structure of D35.
Since the viscous parameter and the stellar mass introduce significant uncertainty, we also performed calculations assuming $\alpha_{\rm vis}=10^{-2},10^{-4}$ and $M_\ast=1.1\,M_\odot, 1.7\,M_\odot$, and adopted the differences from the $\alpha_{\rm vis}=10^{-3}$ case as the uncertainty of the planet mass.
Even though CrA~IRS~2 is a Class I protostar, the gap position and the estimated planet mass are consistent with the Class II sample in \citet{Yamaguchi_2024}, which indicates that the planet mass decreases as the orbital radius increases \citep[see Figure~20 in][]{Yamaguchi_2024}.

\subsubsection{Dust Disk Mass and Gravitational Stability}\label{subsubsec:GI}
To check whether the circumstellar disk around CrA~IRS~2 is gravitationally stable, the Toomre $Q$ parameter \citep[][]{Toomre_1964} is calculated the following formula $Q\equiv c_s\Omega_{\rm K}/\pi G\Sigma_{\rm gas}$, where $c_s$ is the sound speed, $\Omega_{\rm K}$ is the Keplerian angular velocity, $G$ is the gravitational constant, and $\Sigma_{\rm gas}$ is the gas surface density.
For a thin disk, the sound speed can be written as $c_s=h\Omega_{\rm K}$, where $h$ is the pressure scale height, so that the aspect ratio is $h/r$ at radius $r$.
If we approximate the total disk mass within radius $r$ as $M_{\rm disk} \simeq \pi r^2 \Sigma_{\rm gas}$ and assume a constant gas-to-dust mass ratio $\varepsilon=100$, the Toomre $Q$ parameter can be rewritten as
\begin{equation}
   Q = \frac{1}{\varepsilon}\frac{h}{r}\frac{M_\ast}{M_{\rm dust}}, 
   \label{eq:toomreq}
\end{equation}
where $M_\ast$ is the stellar mass and $M_{\rm dust}$ is the dust disk mass within radius $r$.

Using the azimuthally averaged dust surface density profile $\Sigma_{\rm dust}(r)$ derived in this section, we computed the dust mass as $M_{\rm dust}=\int^{R_{\rm 95\%}}_{0}\Sigma_{\rm dust}(r)\, 2\pi r\,dr$, where $R_{95\%}$ is the dust radius (see \S\ref{subsec:morphology}).
Adopting the DSHARP dust opacity $\kappa_\nu = 0.43$ cm$^2$ g$^{-1}$ \citep{Birnstiel_2018} yields a dust mass of $M_{\rm dust} \simeq (3.7 \pm 0.4)\times10^{-3}\,M_\odot$.
Substituting this dust mass and the stellar mass $M_\ast=1.4\pm0.3\,M_\odot$ \citep[][]{Nisini_2005,Fiorellino_2023} into Equation~(\ref{eq:toomreq}), we find that for aspect ratios of $h/r = 0.07$ and 0.14 the Toomre parameter is $Q \simeq 0.18 \pm 0.05$ and $Q \simeq 0.52 \pm 0.13$, respectively, suggesting that the circumstellar disk could be gravitationally unstable.

However, it should be noted that the inferred disk mass (and hence $Q$) is strongly sensitive to the assumed dust opacity.
If instead we adopt the classical dust opacity of \citet{Beckwith_1990}, $\kappa_\nu = 2.3\ {\rm cm^2\,g^{-1}}$, which is about a factor of five larger than the DSHARP value, the inferred dust and gas surface densities (and thus the disk mass) decrease by a comparable factor.
Indeed, once this difference is taken into account, our dust mass becomes $M_{\rm dust} \simeq 9.5\times10^{-4}\,M_\odot$ when using the \citet{Beckwith_1990} opacity, which is comparable to the values reported by \citet{Hsieh_2025} based on the same opacity assumption.
Then, the Toomre $Q$ parameter increases accordingly, so for an aspect ratio of $h/r \simeq 0.14$, we obtain $Q\sim2$, indicating that the disk would be marginally gravitationally stable. 
Given this strong dependence of $Q$ on the uncertain dust opacity and temperature, the current continuum data do not allow us to draw a firm conclusion about whether the disk around CrA~IRS~2 is gravitationally unstable, and the possible role of GI therefore remains uncertain.

\subsection{Possible Role of Interchange Instability}\label{subsec:role_interchange}
We identified that planet–disk interaction with a planet of mass 0.1-1.8\,$M_{\rm Jup}$ is one plausible origin of the outer ring-gap structure in \S\ref{subsec:gap_formation}. 
The planet mass inferred for CrA~IRS~2 is roughly comparable to those inferred for planets of Class II disks in \citet{Andrews_2018} and \citet{Yamaguchi_2024} derived from the gap widths and depths in the same way as \citet{Zhang_2018}.
Given that CrA~IRS~2 is still in the Class~I stage, this similarity suggests that giant planets of comparable mass may form on a much shorter timescale in this system than in those more evolved disks.
These findings suggest that additional physical mechanisms beyond those incorporated into conventional planet-formation models may be required to facilitate sufficiently rapid dust growth and core formation at this early evolutionary stage. 
A possible solution to realize the rapid growth of planet precursors in the younger CrA~IRS~2 system is a drastic magnetic flux removal due to interchange instability as suggested by \citet{Tokuda_2023}. 
In this subsection, we therefore examine how magnetic advection driven by this instability may influence dust growth and planet formation in CrA~IRS~2.

The key to rapid planet formation could be the loss of vertical magnetic flux.
Turbulence driven by MRI is known to enhance radial diffusion and destructive collisions among dust grains, preventing the growth of planetesimals beyond the fragmentation barrier \citep[e.g.,][]{Ida_2008,Nelson_2010}.
However, \citet{Okuzumi_2012} demonstrated that a magnetically inactive region (dead zone) with magnetic flux less than $\sim$10\,mG promotes the dust growth, and planetesimals can grow directly via coagulation by overcoming the fragmentation barrier.

Based on these insights, we suggest a possible scenario in which magnetic flux dissipation due to interchange instability in the CrA~IRS~2 system can suppress MRI-driven turbulence, leading to the formation of an extended dead zone.
\citet{Machida_2025} showed that the magnetic field strength in the inner disk decreases with time due to the interchange instability, and in their simulations it has already declined to $\sim 0.1$\,G in the central region by $\simeq 3\times 10^4$\,yr after protostar formation. 
Given the bolometric temperature of CrA~IRS~2 ($T_{\rm bol}=235$\,K), its age is estimated to be $\sim 0.2$-$0.4$\,Myr, implying that there is ample time for the magnetic field strength in the inner region to be reduced below $\sim 10$\,mG.

Such a weakly magnetized, low-turbulence environment may have created favorable conditions for efficient dust growth and the early formation of protoplanets.
Recent MHD simulations have shown that the dust grows to reach $\sim$0.1-1.0\,cm in a few thousand years during the early star formation stage \citep[e.g.,][]{Tsukamoto_2021,Tu_2022,Koga_2023}.
In addition, coagulation models of porous icy aggregates and global dust-evolution calculations \citep[e.g.,][]{Okuzumi_2012T,Birnstiel_2016} indicate that at radii of a few to several tens of au the growth timescale from submicron grains to pebble-planetesimal sizes is typically $\sim 10^4$-$10^5$\,yr, which is shorter than or at most comparable to the empirical lifetimes of the Class 0/I phases.
Therefore, if interchange-instability-driven magnetic flux dissipation in CrA~IRS~2 indeed leads to the formation of an extended dead zone, substantial dust growth and the onset of planet formation could occur during the Class~I stage, which may help explain the observed ring-gap structure.

However, \citet{Delage_2023} indicated that dust coagulation and settling can eventually increase the gas ionization degree in the disk, leading to the reactivation of MRI turbulence and the shrinkage of the dead zone over time.
Considering the gap structure caused by planet-disk interaction (see \S\ref{subsec:gap_formation}), the CrA~IRS~2 system could be an example in which a protoplanet is formed before MRI turbulence became fully active.
To verify scenarios of rapid planetesimal formation, additional line observations to detect gas kinematics and constrain MRI activity, and non-ideal MHD simulations coupled with dust evolution models, are required.

\section{Summary}\label{sec:summary}
We applied \texttt{PRIISM} imaging to ALMA Band 6 continuum data to analyze the circumstellar disk around the Class I protostar CrA~IRS~2 and provided the dust continuum image of 1.5 times higher spatial resolution than standard imaging.
We found that CrA~IRS~2 hosts the youngest circumstellar disk exhibiting multiple ring structures.
To examine planet-disk interaction as one possible cause of the substructure formation, we compared the depth and width of the outer ring-gap structure measured from the PRIISM image with planet-disk interaction models. 
The formation of the substructures revealed in the PRIISM image can be explained by planet-interaction models involving a protoplanet with a mass of 0.1-1.8\,$M_{\rm Jup}$.
The magnetic flux advection within the disk, driven by the interchange instability, could suppress MRI-driven turbulence and create a favorable environment for the formation of planetesimals in the CrA~IRS~2 system.
This study thus provides new insight into the conditions under which planet formation can occur at the early stages of disk evolution.

\begin{acknowledgments}
The authors appreciate the anonymous referee for all of the comments and advice that helped improve the manuscript and the contents of this study.
The authors thank Dr. S. Okuzumi, Dr. R. Tominaga, Dr. A. Kataoka, and Mr. T. Simokawa for valuable discussions.
This work was supported by a NAOJ ALMA Scientific Research grant (No.2022-22B; MNM), by JSPS KAKENHI 25KJ1947 (AS), 25KJ1921 (MO), 25K07369 (MNM), 23K20238 (K. Tachihara), 20H05645, 21H00049, and 21K13962 (K. Tokuda), and by Kagawa University Research Promotion Program 2025 Grant Number 25K0D015 (K. Tokuda).
This paper makes use of the following ALMA data: ADS/JAO.ALMA\#2019.1.01792.S ALMA is a partnership of ESO (representing its member states), NSF (USA) and NINS (Japan), together with NRC (Canada), MOST and ASIAA (Taiwan), and KASI (Republic of Korea), in cooperation with the Republic of Chile. 
The Joint ALMA Observatory is operated by ESO, AUI/NRAO and NAOJ.
The National Radio Astronomy Observatory is a facility of the National Science Foundation operated under cooperative agreement by Associated Universities, Inc.
Data analysis was in part carried out on a common-use data analysis computer system at the Astronomy Data Center, ADC, of the National Astronomical Observatory of Japan.
\end{acknowledgments}

\facility{ALMA}

\software{astropy \citep[e.g.,][]{astropy_2022},
          CASA \citep[][]{CASA_2022},
          emcee \citep[][]{Foreman-Mackey_2013},
          matplotlib \citep[][]{Hunter_2007},
          PRIISM \citep[][]{Nakazato_2020,Nakazato_2020b},
          protomidpy \citep[][]{Aizawa_2024},
          SciPy \citep[][]{Virtanen_2020},
}

\appendix
\restartappendixnumbering
\section{Conservative Determination of the TSV Parameter}\label{sec:tsv_setting}

\begin{figure*}[t]
    \centering
    \includegraphics[width=0.80\linewidth]{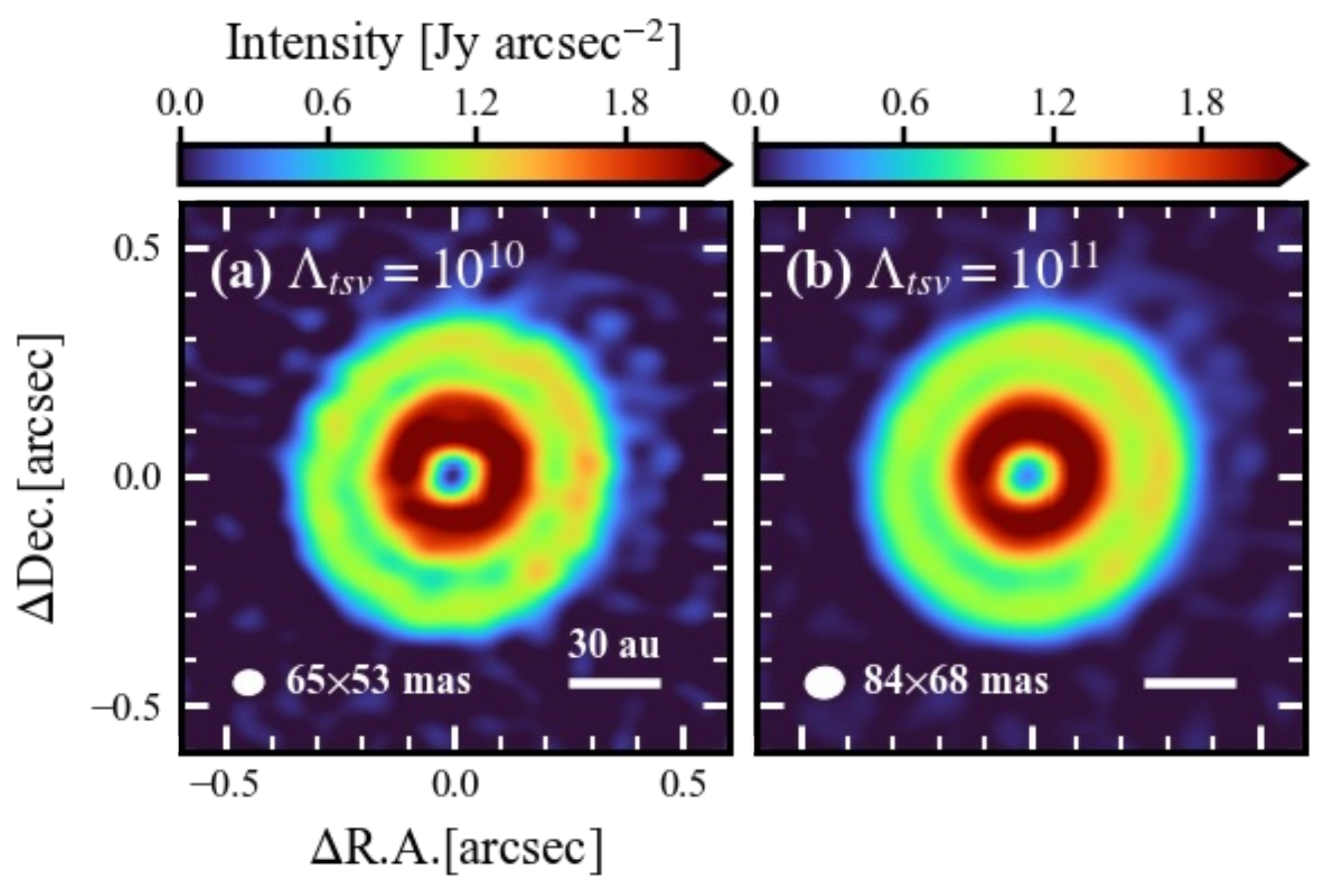}
    \caption{ALMA Band 6 (1.3\,mm) dust continuum images obtained using \texttt{PRIISM}.
             (a) Image produced with the hyperparameters $(\Lambda_l, \Lambda_{tsv}) = (10^4, 10^{10})$.
             (b) Same as panel (a), but with $(\Lambda_l, \Lambda_{tsv}) = (10^4, 10^{11})$.
             Both images share the same color scale.
             The white ellipse in each panel indicates the effective spatial resolution $\theta_{\rm eff}$ for each image.
             }
    \label{fig:tsv_cont}
\end{figure*}

\begin{figure}
    \centering
    \includegraphics[width=\linewidth]{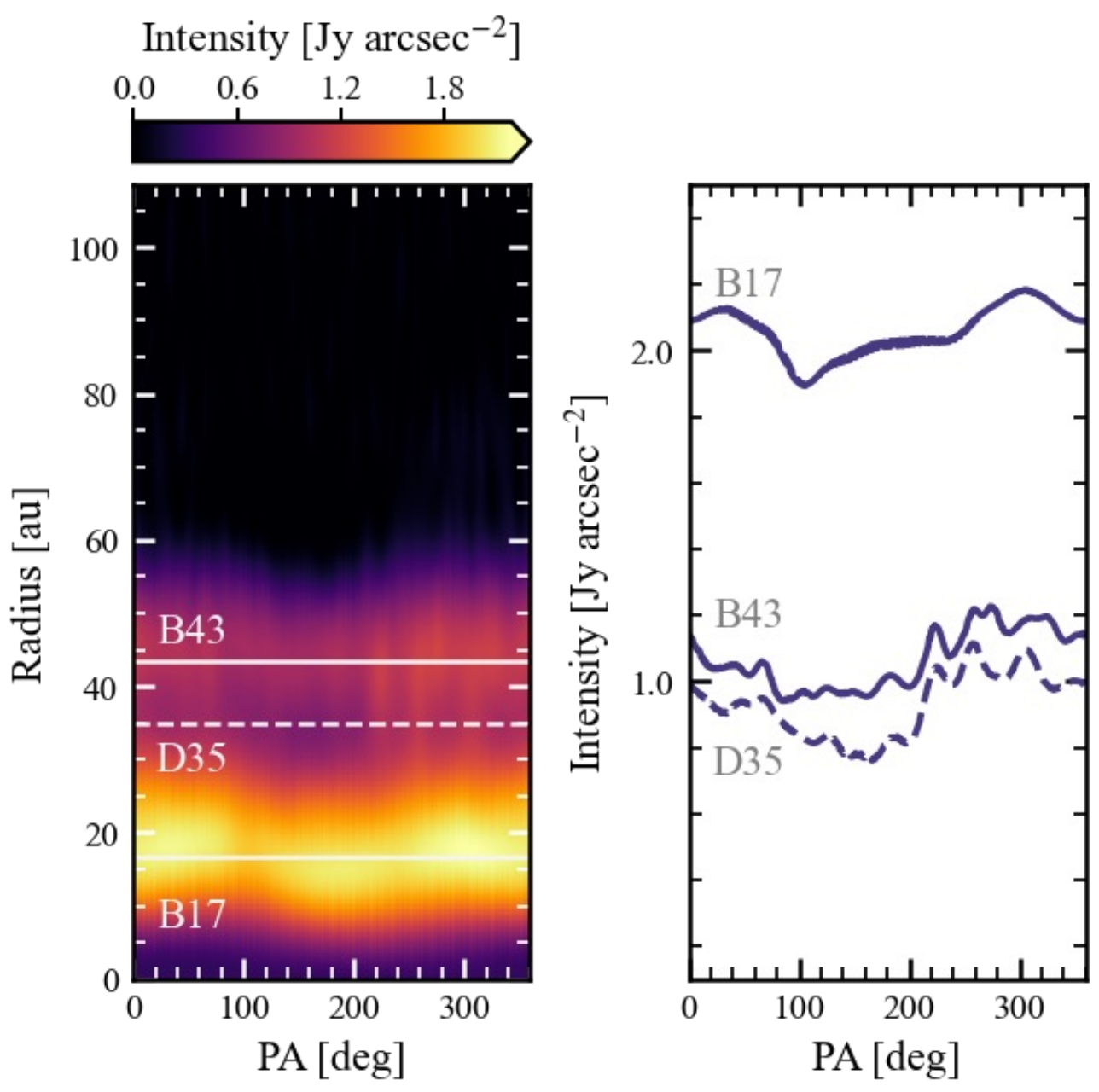}
    \caption{(Left) The PRIISM image with $(\Lambda_l, \Lambda_{\rm tsv}) = (10^4, 10^{11})$, where the brightness distribution has been deprojected using $i_{\rm disk}$=21.2$^\circ$ and PA=145.3$^\circ$ and then transformed into a polar-coordinate map.
    (Right) Intensity profile in the azimuthal direction at the inner and outer rings (B17 and B43), and the gap structure (D35).
    }
    \label{fig:paprofile}
\end{figure}

\begin{figure}
    \centering
    \includegraphics[width=\linewidth]{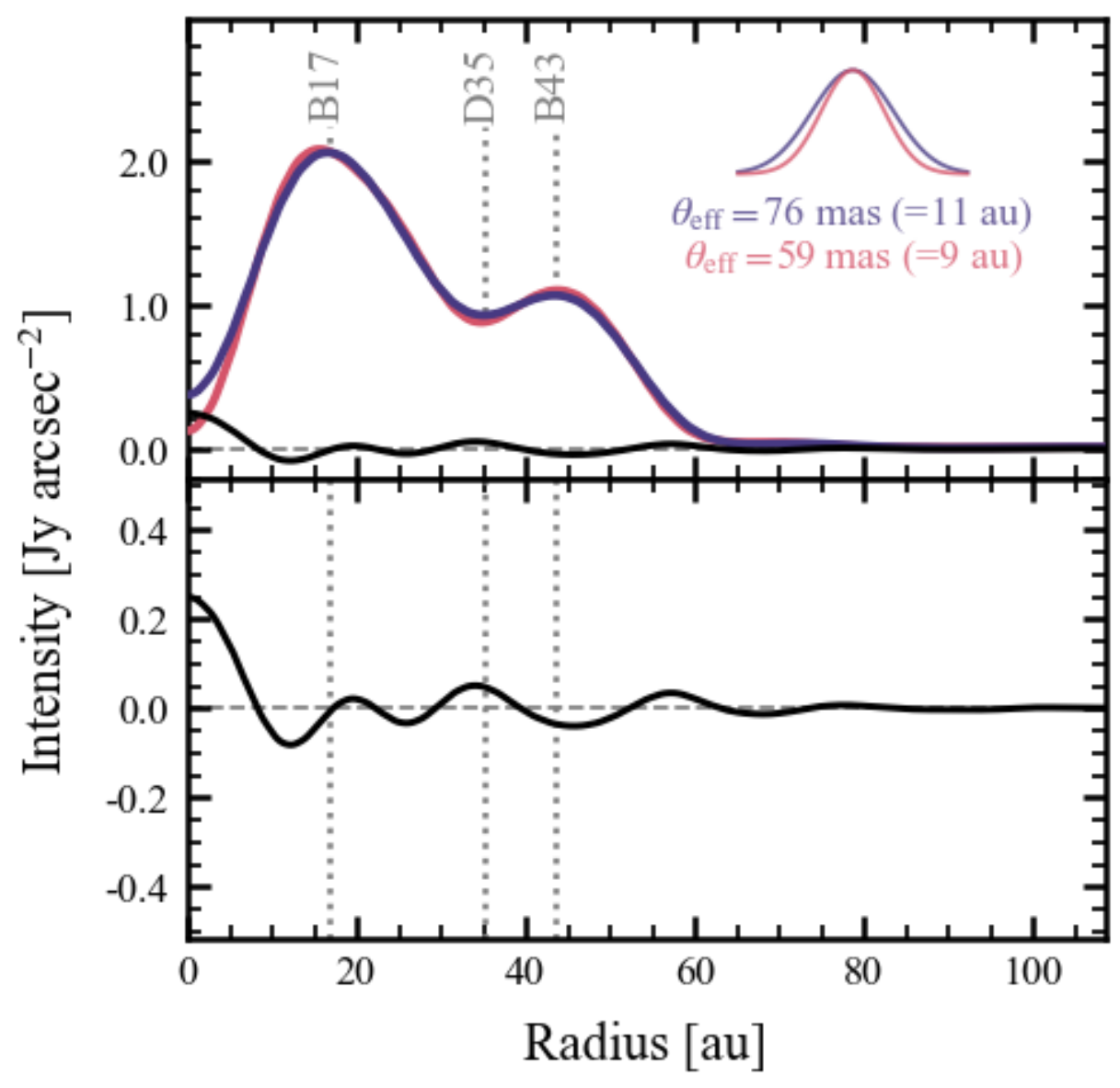}
    \caption{(Top) Radial intensity profile averaged over the full azimuthal angle.
             The violet solid curve shows the profile of the PRIISM image with $(\Lambda_l, \Lambda_{tsv}) = (10^4, 10^{11})$, and the red one shows that with $(\Lambda_l, \Lambda_{tsv}) = (10^4, 10^{10})$.
             The black curve shows the residual between the two profiles.
             (Bottom) Same as the top panel, but only the residual profile in a different intensity range.}
    \label{fig:tsv_prof}
\end{figure}

Figure~\ref{fig:tsv_cont}(a) shows the PRIISM image reconstructed with the parameters $(\Lambda_l, \Lambda_{tsv}) = (10^4, 10^{10})$, which were selected using the 10-fold CV method.
The image exhibits clumpy brightness distributions in the outer ring.
These features are likely artificial because the imaging process prioritizes fitting the bright and compact inner ring, which can also induce compact patchy structures in the extended region surrounding it \citep[][]{Yamaguchi_2024}.
In particular, the integration time was less than one minute, resulting in the incomplete (sparse) sampling of the observed visibility, which in turn leads to poor sensitivity.
Such clumps can affect the characterization of the disk structure in two ways.
First, they can induce fluctuations in the azimuthal peak positions used for elliptical fitting, leading to larger uncertainties in the derived inclination and position angle of the outer ring.
Second, they can distort the azimuthally averaged radial intensity profile, potentially leading to inaccurate estimates of the gap width and depth.
Therefore, we conservatively select the image with $\Lambda_{tsv}=10^{11}$, as shown in Figure~\ref{fig:tsv_cont}(b), which is one order of magnitude larger than the value selected by the CV method.
The flux density of the image with $\Lambda_{tsv}=10^{11}$ is comparable to that of Figure~\ref{fig:tsv_cont}(a), but suppresses the clump features and shows a more smoothed distribution than Figure~\ref{fig:tsv_cont}(a).

To quantify the azimuthal asymmetry, we transformed the deprojected PRIISM image with $(\Lambda_l, \Lambda_{\rm tsv}) = (10^4, 10^{11})$ into polar coordinates and extracted intensity profiles as a function of azimuth at the radii of the inner ring (B17), gap (D35), and outer ring (B43). 
For each position, we calculated the standard deviation of the azimuthal intensity profiles and expressed it as a fraction of the profile's peak intensity. 
Figure~\ref{fig:paprofile} presents the azimuthal intensity profiles and the residual between them, which are $3.4\%$ at B17, $7.2\%$ at B43, and $8.8\%$ at the gap radius D35 to each peak of the azimuthal intensity profile. 
These $\lesssim 10\%$ variations in the intensity profile indicate that the azimuthal asymmetry is modest.
For comparison, strongly asymmetric transition disks such as HD~142527 exhibit azimuthal intensity contrasts of order unity \citep[e.g.,][]{Fukagawa_2013,Boehler_2017}, much larger than those found in CrA~IRS~2. 
Therefore, the impact of such weak azimuthal asymmetry on the one-dimensional radial intensity profile used in our analysis is expected to be small.

We also compared the two images in terms of the radial intensity profile averaged over the full azimuthal angle. 
Figure~\ref{fig:tsv_prof} shows the profiles derived from the PRIISM images with $(\Lambda_l, \Lambda_{\rm tsv}) = (10^4, 10^{10})$ and $(10^4, 10^{11})$, together with the residual profile between them. 
Although a noticeable difference appears near $r$=0\,au, this is simply due to the different effective angular resolutions $\theta_{\rm eff}$ of the two reconstructions, which affect the width of the peak associated with the inner ring (B17).
As shown in the bottom panel of Figure~\ref{fig:tsv_prof}, the residuals outside this innermost region remain at the level of $\sim 0.05$\,Jy\,arcsec$^{-2}$ (i.e., $\sim 2\%$ of the peak intensity in the PRIISM image with $(\Lambda_l, \Lambda_{\rm tsv}) = (10^4, 10^{11})$), suggesting that the two profiles agree well.
These tests demonstrate that our conservative choices of the PRIISM imaging parameters do not significantly affect the recovered disk morphology or the radial intensity profile, indicating that our analysis based on these profiles is robust against reasonable variations in the imaging setup.

\restartappendixnumbering
\section{Validation of the Effective Spatial Resolution of the PRIISM Image}\label{sec:resolution}

\begin{figure}[t]
    \centering
    \includegraphics[width=\linewidth]{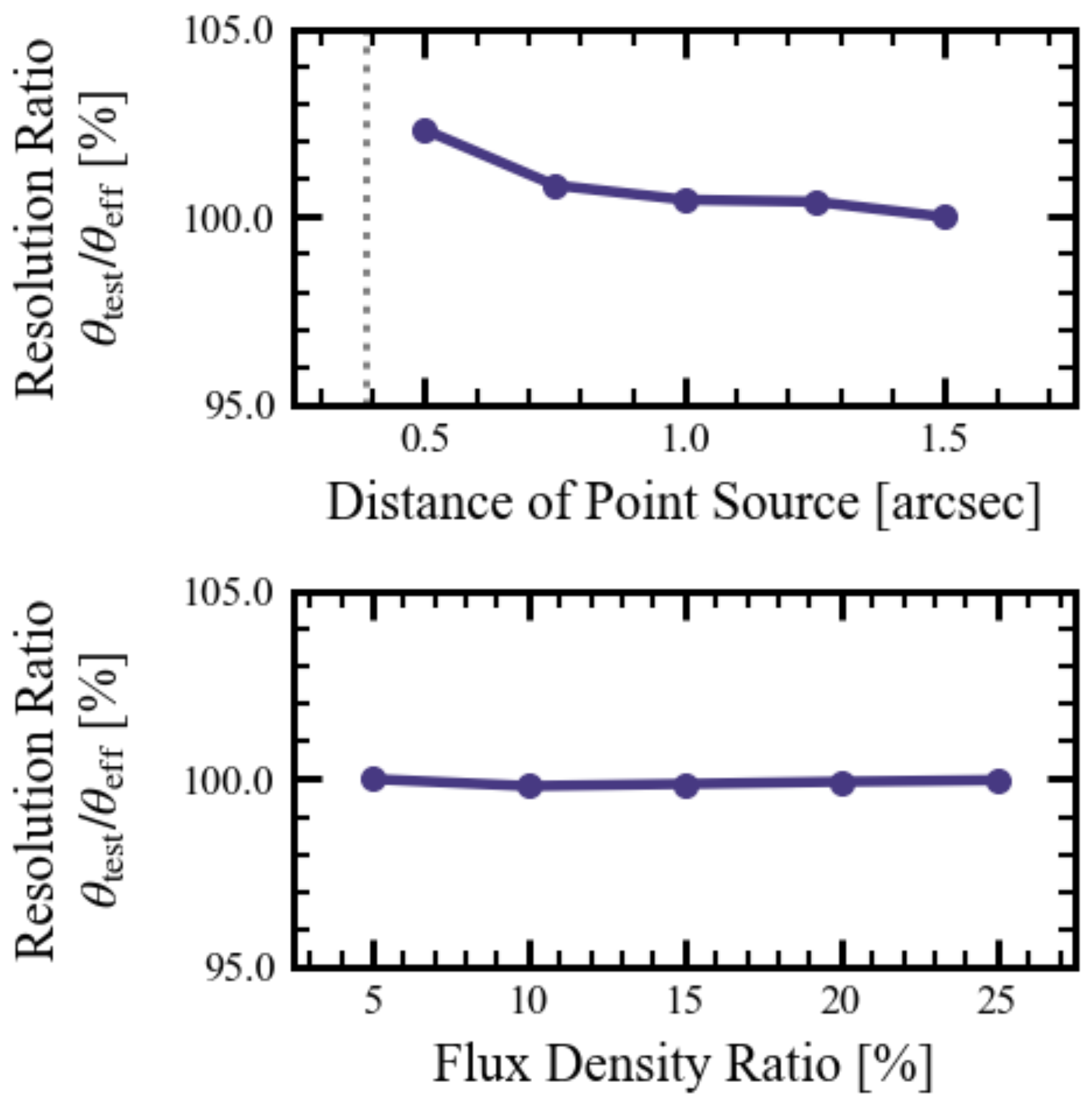}
    \caption{Relationships between the resolution ratio relative to the effective spatial resolution, $\theta_{\rm eff}$, of the PRIISM image (Figure~\ref{fig:continuum}b).
    (Top) The distance from the disk center to the injected point source.
    The gray vertical dotted line indicates the dust disk radius, $R_{95\%}=57.8$\,au ($\sim0\farcs39$).
    (Bottom) The total flux density ratio of the point source relative to the circumstellar disk.
    In both panels, the effective resolution, $\theta_{\rm eff}$, was derived using an artificial point source with 5\% of the total flux density, placed at a distance of $1\farcs5$ from the disk center.
    }
    \label{fig:resolution}
\end{figure}

We evaluated the effective spatial resolution ($\theta_{\rm eff}$) of the PRIISM image shown in Figure~\ref{fig:continuum}(b) using the point-injection method.  
In this approach, an artificial point source corresponding to 5\% of the total flux density ($F_\nu$) of the disk was added at a distance of $1\farcs5$ from the disk center.  
This point source was used as a reference to measure the effective spatial resolution of the reconstructed image.  
To estimate the uncertainty of $\theta_{\rm eff}$, we performed a series of tests by injecting several artificial point sources with different positions and flux densities into images reconstructed with the same hyperparameters $\left(\Lambda_l,\Lambda_{tsv}\right)=\left(10^4,10^{11}\right)$ as those adopted in Figure~\ref{fig:continuum}(b).  
We evaluate the stability of the point source injection method by varying the position and flux density of the artificial point source. 

The top panel of Figure~\ref{fig:resolution} shows how $\theta_{\rm test}$ varies with the distance of the injected point source from the disk center.  
The total fluxes of the point sources were fixed to be 5\% of $F_\nu$.
The measured $\theta_{\rm test}/\theta_{\rm eff}$ ratios are nearly constant across all tested positions, remaining within 1\% of variation except for the source placed at $0\farcs5$.  
A small increase of up to 3\% was found at this position, likely due to contamination from faint noise emission around the circumstellar disk.  
Nevertheless, this difference is within the measurement uncertainty, confirming that $\theta_{\rm eff}$ remains robust and position-independent across the image. 
As a practical consideration, the injected point source was placed in an emission-free region at a projected distance from the phase center.

Figure~\ref{fig:resolution} shows the dependence of the effective spatial resolution on the flux density. 
We examine how the resolution changes when increasing the total flux of point sources from 5\% to 25\%, while keeping them fixed at a distance of $1\farcs5$ from the disk center.
In this test, point sources with flux ratios ranging from 5\% to 25\% of the total disk flux were placed at a fixed distance of $1\farcs5$ from the disk center.  
The measured $\theta_{\rm test}/\theta_{\rm eff}$ ratios remain consistently close to 100\% across all flux densities, indicating that the point-injection method is likely unaffected by the signal strength of the injected source.

In summary, our evaluation confirmed that $\theta_{\rm eff}$ is stable against variations in both the source position and the flux.  
In addition, both tests demonstrate that the point-injection method provides a reliable and reproducible estimate of the effective spatial resolution $\theta_{\rm eff}$ in the PRIISM image.  

\bibliography{reference}{}
\bibliographystyle{aasjournal}

\end{document}